\documentclass[twocolumn, 10pt, superscriptaddress,longbibliography, nofootinbib, pra]{revtex4-2}
\usepackage[english]{babel}

\usepackage{graphicx}
\usepackage{setspace}
\usepackage{amsmath,amssymb,amsthm}
\usepackage{bbold}
\usepackage[colorlinks=true, linkcolor=blue, citecolor=gray]{hyperref}
\usepackage[dvipsnames,svgnames]{xcolor}
\usepackage{enumitem}
\usepackage[utf8]{inputenc}
\usepackage{cancel} 
\usepackage{qcircuit}
\usepackage[ruled,vlined]{algorithm2e}
\usepackage[version=4]{mhchem}
\usepackage{siunitx}
\usepackage{physics}
\usepackage{ltablex}
\usepackage{subfigure}
\usepackage{tabularx}
\usepackage{array}
\usepackage{comment}

\definecolor{LEI-orange}{cmyk}{0,.62,.97,0} 
\newcommand{\topic}[1]{
	\textcolor{LEI-orange}{\emph{#1} ---}
}
\renewcommand{\topic}[1]{}

\AtBeginDocument{\RenewCommandCopy\qty\SI} 

\DeclareSIUnit\angstrom{\text {Å}}

\newcommand{\Leiden}{\affiliation{%
Instituut-Lorentz, Universiteit Leiden, 2300RA Leiden, The Netherlands}}

\newcommand{\VU}{\affiliation{%
Theoretical Chemistry, Vrije Universiteit, 1081HV Amsterdam,
The Netherlands}}

\begin{document}
\title{FragPT2:\\
Multi-Fragment Wavefunction Embedding with Perturbative Interactions}

\author{Emiel Koridon}
\email{koridon@lorentz.leidenuniv.nl}
\Leiden \VU

\author{Souloke Sen}
\Leiden \VU

\author{Lucas Visscher}
\VU

\author{Stefano Polla}
\Leiden

\date{\today}

\begin{abstract}
    Embedding techniques allow the efficient description of correlations within localized fragments of large molecular systems, while accounting for their environment at a lower level of theory.
    We introduce FragPT2: a novel embedding framework that addresses multiple interacting active fragments.
    Fragments are assigned separate active spaces, constructed by localizing canonical molecular orbitals.
    Each fragment is then solved with a multi-reference method, self-consistently embedded in the mean field from other fragments.
    Finally, inter-fragment correlations are reintroduced through multi-reference perturbation theory.
    Our framework provides an exhaustive classification of inter-fragment interaction terms, offering a tool to analyze the relative importance of various processes such as dispersion, charge transfer, and spin exchange.
    We benchmark FragPT2 on challenging test systems, including \ce{N_2} dimers, multiple aromatic dimers, and butadiene. 
    We demonstrate that our method can be succesful even for fragments defined by cutting through a covalent bond.
\end{abstract}

\maketitle


%
\section{Introduction}
\label{sec:introduction}
\topic{MC methods}
Multi-configurational (MC) wavefunction-based methods have long been the workhorse of ab-initio quantum chemistry, particularly for systems with low-lying or degenerate electronic states \cite{park2020multireference,szalay2012multiconfiguration}.
Practical MC approaches, such as the complete active space self-consistent field (CASSCF) \cite{roos2007complete}, require defining an active space comprising a subset of the most chemically relevant orbitals.
Within this space, electron correlations are calculated exactly by a configuration interaction (CI) wavefunction, a superposition of all electronic configurations formed from a given set of active electrons and orbitals. 
The number of these configurations scales exponentially with the size of the active space, limiting the application of these methods to small systems.
There have been substantial efforts to expand the size of the active space: some try to restrict the number of excitations by partitioning the active space \cite{olsen1988determinant,malmqvist1990restricted,fleig2003generalized,ma2011generalized, li2011strong,li2013splitgas}, others involve adaptive procedure to select the configurations with the largest weights \cite{smith2017cheap,levine2020casscf}.
Radically different approaches to constructing a compressed CI wavefunction include tensor-network algorithms such as the density matrix renormalization group (DMRG) \cite{baiardi2020density}, quantum monte carlo (QMC) methods \cite{austin2012quantum}, or various kinds of quantum algorithms \cite{bauer2020}.

\topic{Fragmentation and embedding}
A more pragmatic approach for extending multi-configurational computations to larger systems relies on the concepts of \emph{fragmentation} and \emph{embedding} \cite{gordon2012fragmentation,collins2015energy,raghavachari2015accurate,fedorov2011geometry}.
Fragmentation exploits the inherent locality of the problem, describing a system as a composition of simpler subsystems.
Each subsystem is then treated with a higher level of theory. The subsystems are then recombined by embedding them in each other's environment at a lower level of theory. 
The subsystem orbitals can be constructed in various ways, with the most prominent method being Density Matrix Embedding Theory (DMET) \cite{knizia2012density,knizia2013density,bulik2014density, yalouz2022quantum}. 
DMET constructs fragment and bath orbitals based on the Schmidt decomposition of a trial low-level (eg. Hartree-Fock) single-determinant wavefunction of the full system. 
A high-level calculation (e.g.~FCI, Coupled-Cluster \cite{bulik2014electron,wouters2016practical}, CASSCF \cite{pham2018can}, DMRG \cite{wouters2016practical,zheng2016ground,chen2014intermediate} or auxiliary-field QMC \cite{zheng2017cluster}) is then performed on the fragment orbitals.
Subsequently, the low-level wavefunction is fine-tuned self-consistently via the introduction of a local correlation potential.
Fragmentation and embedding have also been studied in the context of DFT \cite{jones2020embedding, sen2023computational}.
MC wavefunction-based methods that explicitly construct localized active spaces for each fragment include the Active Space Decomposition method~\cite{parker2013communication},  cluster Mean Field (cMF)~\cite{jimenez2015cluster} and Localized Active Space Self-Consistent Field (LASSCF)~\cite{hermes2019multiconfigurational,hermes2020variational}.


\topic{Correlations between fragments and environment}
While fragmentation methods have shown success in reducing the complexity in treating localized static correlations, they typically don't capture inter-fragment correlations.
Especially weak, dynamical, correlations between the different fragments and between fragments and their environment
can be crucial for obtaining an accurate description of the full system \cite{Rummel2024}. 
In CAS methods, the fragment-environment correlations can be retrieved using Multi-Reference Perturbation Theory (MRPT) \cite{schrodinger1926quantisierung} methods like Complete Active Space Second-Order Perturbation Theory (CASPT2) \cite{andersson1990secondorder} and N-Electron Valence Second-Order Perturbation Theory (NEVPT2) \cite{angeli2001introduction,angeli2002n}.
Some methods have been developed to also recover inter-fragment correlations in embedding schemes either variationally \cite{abraham2020selected}, perturbatively \cite{jimenez2015cluster,hapka2021symmetry,papastathopoulos2022coupled,xu2013block}, or via a coupled-cluster approach \cite{wang2020describing}. 
Although treating strong correlations between fragments remains challenging, there has been some work in this direction \cite{he2020zeroth,coughtrie2018embedded}. 
In the field of quantum algorithms, a recent work proposed to treat inter-fragment entanglement with a Unitary Coupled Cluster ansatz using the LASSCF framework \cite{otten2022localized}.

\topic{Our method}
In this work, we introduce and benchmark a novel active space embedding framework, which we call FragPT2.
Based on a user-defined choice of two molecular fragments (defined as a partition of the atoms in the molecule), we employ a top-down localization scheme that generates an orthonormal set of localized molecular orbitals, ordered by quasi-energies and assigned to a specific fragment.
Using these localized orbitals, we define separate and orthogonal fragment active spaces.
Our orbital fragmentation scheme is straightforward, it does not require iterative optimization, and it allows to define fragment orbitals even when the fragments are covalently bonded; on the downside, a good choice of fragments based on chemical intuition is crucial for the success of our method.
Within each fragment's active space, we self-consistently find the MC ground state influenced by the mean field of the other fragment (defined as a function of the fragment 1-particle reduced density matrix).

\topic{Product state wavefunctions}
The factorized state obtained with our method has a similar structure to the wavefunction used in LASSCF and cMF, as these methods also construct product state wavefunctions of MC states defined on fragmented active spaces.
The cMF method is designed for the 1D and 2D Fermi-Hubbard model.
It is based on expressing the ground state wavefunction as a tensor product of many-body states defined on local fragments.
The fragment orbitals are then optimized self-consistently to minimize the total energy of the considered product state.
Inter-fragment correlations are then recovered in second-order perturbation theory, using excited fragment eigenstates as perturbing functions.
On the other hand, LASSCF exploits a modified DMET algorithm to construct fragments.
Starting from a product state, a Schmidt decomposition is used to define fragment and bath orbitals for each fragment.
Similarly to cMF, the product state and fragment definition are then optimized self-consistently.
The resulting method can be made fully variational with respect to both CI and orbital coefficients \cite{hermes2020variational}.
%
In contrast, in our approach, active fragment orbitals are defined in top-down fashion, starting from a set of reference canonical molecular orbitals.
Our method is variational with respect to the considered (fragment CI) parameters, and does not require any orbital optimization.
As a trade-off for the simplicity of the method, we expect our product wavefunction to have a higher energy than the orbital-optimized LASSCF for the same fragment active space sizes.
We instead aim to recover the remaining inter-fragment correlations perturbatively.


\topic{FragPT2}
To this end, our product state will be used as a starting point for MRPT to recover inter-fragment correlations.
The interactions between fragments can be naturally classified on the basis of charge and spin symmetries imposed on the single fragments, offering analytic insight into the nature of these correlations.
Differently from cMF, the perturbing functions are chosen on the basis of electronic excitation operators present in the original electronic Hamiltonian, and organized according to a partially contracted basis akin to MRPT methods like PC-NEVPT2 \cite{angeli2001introduction,angeli2002n}.
We apply our method to challenging covalently and non-covalently bonded fragments with moderate to strong correlation, providing qualitative estimates of the contributions from various perturbations to the total correlation energy within the active space.

\topic{In this paper}
The rest of this paper is organized as follows: in Section~\ref{sec:method} we detail our FragPT2 algorithm for multi-reference fragment embedding. 
In Section~\ref{sec:numerics}, we perform numerical tests of the method on a range of challenging chemical systems, ranging from the non-covalently bonded but strongly correlated \ce{N2} dimer to covalently bonded aromatic dimers and the butadiene molecule. 
In Section~\ref{sec:outlook} we present an outlook on future research directions, proposing possible improvements for the method and an application in the field of FragPT2 in the field of variational quantum algorithms.
Finally, in Section~\ref{sec:conclusion} we give concluding remarks.

\begin{figure}[t]
\centering
\includegraphics[width=\columnwidth]{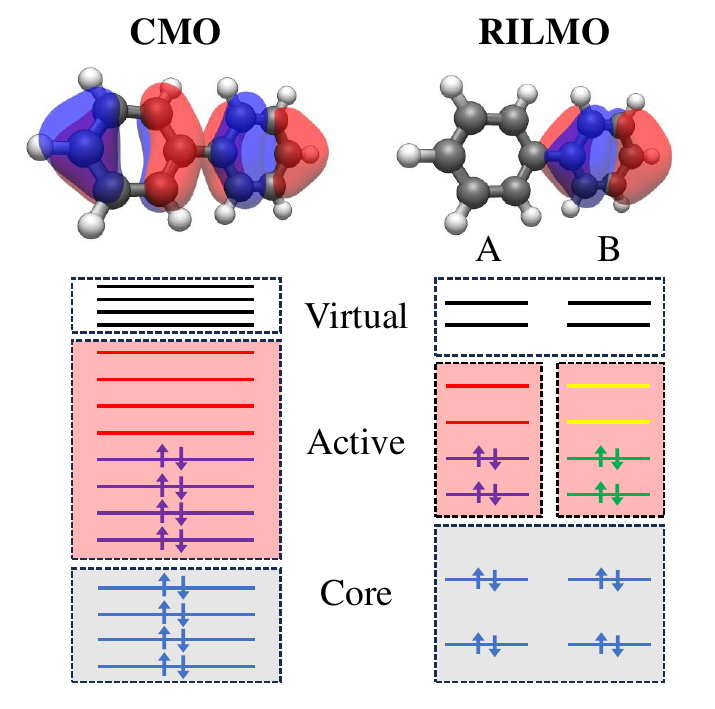}
\caption{
    \textbf{Example of fragmentation and definition of the fragment active spaces}.
    \textbf{(Left)} Active space selection for the entire biphenyl molecule.
    The CAS treatment separates the canonical molecular orbitals (CMOs) based on their energy ordering, obtaining a set of doubly-occupied core orbitals, a set of empty virtual orbitals, and a set of active orbitals around Fermi energy used to describe correlations. 
    We illustrate the highest occupied molecular orbital. 
    \textbf{(Right)} Fragment active space selection for the left and right fragments of the biphenyl molecule.
    After the localization procedure, we obtain Recanonicalized Intrinsic Localized Molecular Orbitals (RILMOs), where the orbitals are assigned to either fragment A or B.
    We can still select core, active and virtual orbitals for each fragment based on an approximate energy ordering, obtained through the recanonicalization procedure. 
    Here we depict the highest occupied RILMO for the right fragment.
    Using our method, we can half the size of the required active space since the multi-reference solver is applied to just one fragment at a time. 
    The correlations between the localized active spaces can be retrieved afterwards with perturbation theory.
}\label{fig:active_space}
\end{figure}

\section{FragPT2 method}
\label{sec:method}

In this section, we introduce a novel method for fragmented multi-reference calculations with perturbative corrections: FragPT2.
This method works by dividing the active space of a molecule into localized subspaces that can be treated separately using a MC solver, as illustrated in Figure~\ref{fig:active_space}.
The cost of MC methods scales quickly with the size of the treated active space (e.g.~exponentially in the case of FCI); splitting the system into smaller active spaces allows the treatment of larger systems for an affordable computational cost.
In this work, we focus on the special case of two active fragments called $A$ and $B$; however, our method can be promptly generalized to the multiple fragment case as discussed in Section~\ref{sec:extensions}.
Our method requires the user to define the molecular fragments as an input.
The choice of fragmentation should be based on chemical intuition, aiming at minimizing inter-fragment correlations; a good choice is crucial to the success of the method.
Our method allows to recover some inter-fragment correlations, allowing fragmentations that \emph{break a covalent bond} (like the one shown in Figure~\ref{fig:active_space} for biphenyl), i.e.~where two atoms on either side of a covalent bond are assigned to different fragments.
The number of bonds broken in fragmentation should, however, be kept to a minimum.

First, in Sec~\ref{sec:orbitals} we introduce the construction of the localized orbitals and the definition of the fragment active spaces. 
In Section~\ref{sec:ham_decomp} we define fragment Hamiltonians by embedding each fragment in the mean field of the other.
Applying separate MC solvers to each fragment Hamiltonian, we show how to obtain a fragment product state $\ket{\Psi_0}$ which will be the reference state for subsequent perturbative expansions.
Finally, in Section~\ref{sec:pt2} we decompose the full Hamiltonian into a sum of the solved fragment Hamiltonians and a number of inter-fragment interaction terms. 
We classify these terms on the basis of fragment symmetries and describe a method to treat them in second-order perturbation theory.

\subsection{Construction of re-canonicalized intrinsic localized molecular orbitals}
\label{sec:orbitals}

In order to define the fragment subspaces, we follow the top-down procedure introduced in references~\citenum{senjean2021generalization}~and~\citenum{sen2023characterization}, based on localizing pre-computed molecular orbitals.
First, we calculate a set of canonical molecular orbitals (CMOs) for the whole system (other choices for molecular orbitals are discussed in Section~\ref{sec:conclusion}). Distinct Hartree-Fock calculations are also run on each fragment, capped if necessary to saturate bonds severed in the fragmentation.
We then choose a valence space, removing a set of hard-core and hard-virtual orbitals far from Fermi energy in both the supermolecular and the fragment calculations. The remaining valence fragment orbitals define the target localized active spaces and are called reference fragment orbitals (RFOs). These RFOs are non-orthogonal and only serve to depolarize the valence CMOs, providing an orthonormal set of intrinsic fragment orbitals (IFOs) of the same dimension as the RFO basis. These IFOs are expressed in the CMO basis and could already be assigned to a particular fragment. They do however mix occupied and virtual spaces and we therefore merely use them to define the localization function in Pipek-Mezey localization~\cite{pipek1989fast} of the CMOs. After recanonicalization (block-diagonalizing the Fock matrix within each fragment), we obtain a set of Recanonicalized Intrinsic Localized Molecular Orbitals (RILMOs), partitioned in fragment subspaces, that together span exactly the occupied space of the original CMOs~\cite{senjean2021generalization} plus the chemically relevant valence virtual space. The active spaces for each fragment are illustrated in Figure~\ref{fig:active_space}.

In this work we also consider covalently bonded fragments, where there is an ambiguity in assigning one occupied orbital representing the inter-fragment bond to either fragment. The same ambiguity holds for one unoccupied antibonding orbital, which can also be assigned to either fragment.
To eliminate this arbitrariness, we introduce a bias so that any such (anti-)bond is always assigned to the first fragment.
This enables us to define a natural fragmentation for covalently bonded dimer molecules.
As noted already above, in order to generate the required IFO basis for this calculation, we need to deal with ``dangling'' bonds that are severed in the fragmentation process. For each fragment we simply saturate these by adding a hydrogen atom to the fragment. The thus produced fragment orbitals are well suited as RFOs, but do yield one additional orbital in the span of the RFOs and IFOs. Accepting this feature, the ROSE code reported in reference \citenum{senjean2021generalization} could be used without modification. In a forthcoming paper, we plan to discuss the localization of higher lying virtuals for which the RILMO generation does need to be modified (see also reference \citenum{sen2023characterization} for non-covalently bonded subsystems) by removing the capping basis from the RFO space. For the covalently bonded dimer systems tested in this work, the unmodified RILMO generation could be used with only a bias in the selection procedure to assign both the bond and the antibond to the same fragment.

\subsection{Fragment embedding}\label{sec:ham_decomp}

\topic{total active space Hamiltonian}
The total Hamiltonian in the combined active space spanned by both fragments is given by
\begin{align}\label{eq:ham_tot}
    H = \sum_{pq\in A\cup B}
    h_{pq} E_{pq} + \frac{1}{2}\sum_{pqrs\in A\cup B}
    g_{pqrs} e_{pqrs},
\end{align}
where we use the spin-adapted excitation operators 
\begin{equation}
\begin{aligned}
    E_{pq} &= \sum_\sigma a_{p\sigma}^\dagger a_{q\sigma}, 
    \\
    e_{pqrs} &= \sum_{\sigma \tau} a_{p\sigma}^\dagger a_{r\tau}^\dagger a_{s\tau} a_{q\sigma} = E_{pq}E_{rs} - \delta_{qr}E_{ps}.
\end{aligned}
\end{equation}
This Hamiltonian includes all interactions of all active orbitals.
Our embedding scheme aims at decomposing this Hamiltonian as $H = H^0 + H'$, where $H^0$ includes intra-fragment terms and a mean-field inter-fragment term, and can be solved exactly with separate in-fragment MC solvers.  
The residual inter-fragment interactions $H'$ are treated separately with perturbation theory, as described in Section~\ref{sec:pt2}.

\topic{Fragment product state and local symmetries}
To facilitate the use of separate MC solvers for each fragment, we constrain the wavefunction of the total system to be a product state over the two fragments,
\begin{align} \label{eq:prod-state}
    \ket{\Psi^0} = \ket{\Psi_A} \ket{\Psi_B},
\end{align}
where $\ket{\Psi_X}$ is a many-body wavefunction in the active space of fragment $X$, similar in spirit to cMF and LASSCF.
We further restrict each fragment wavefunction $\ket{\Psi_X}$ to have fixed, integer charge and spin.
Note that the conservation of spin and charge on each fragment is not a symmetry of the subsystem; however, this assumption is crucial to construct separate efficient MC solvers.
Inter-fragment charge transfer and spin exchange processes are later treated in perturbation theory.

\topic{Restricted Hamiltonian}
Under these constraints, we can simplify the expression of $H$ by removing all the terms that do not respect charge and spin conservation on each fragment separately (as their expectation value of $\ket{\Psi^0}$ would anyway be zero).
The remaining Hamiltonian can be then decomposed as $H_A + H_B + H_{AB}$, with terms
\begin{equation}\label{ham_local}
    H_X = \sum_{pq\in X} h_{pq} E_{pq} + \frac{1}{2}\sum_{pqrs \in X} g_{pqrs} e_{pqrs}
\end{equation}
(with $X \in \{A, B\}$), that only act non-trivially on a single fragment, and a term
\begin{equation}\label{eq:ham_ab}
    H_{AB} = \sum_{pq\in A}\sum_{rs\in B} g^\prime_{pqrs} E_{pq}E_{rs}
\end{equation}
(where $g^\prime_{pqrs} = g_{pqrs} - \frac{1}{2}g_{psrq}$), that includes interactions preserving local spin and charge.
The term $H_{AB}$ still introduces inter-fragment correlations; one way to make the fragments completely independent would be to also treat this term perturbatively (this is the choice made in SAPT \cite{hapka2021symmetry}). 
However, including an effective mean-field interaction (originating from $H_{AB}$) in the non-perturbative solution improves the quality of our $\ket{\Psi^0}$.

\topic{Embedding Hamiltonian $H^0$, mean field} 
To construct the effective Hamiltonian $H^\text{eff}_X$ for each fragment we use a mean-field decoupling approach. 
We write the excitation operator as its mean added to a variation upon the mean: $E_{pq} = \expval{E_{pq}} + \delta E_{pq}$.
The mean is just the one-particle reduced density matrix (1-RDM) of one of the fragments, $\gamma^{X}_{pq} = \mel{\Psi_X}{E_{pq}}{\Psi_X}$. 
By substituting in Eq.~\eqref{eq:ham_ab} we obtain
\begin{align}
    \sum_{pq\in A}\sum_{rs\in B} g^{\prime}_{pqrs} \left[E_{pq}\gamma^B_{rs} + \gamma^A_{pq}E_{rs} - \gamma^A_{pq} \gamma^B_{rs} + \delta E_{pq} \delta E_{rs} \right].
\end{align} 
The term $\delta E_{pq} \delta E_{rs}$ will necessarily have zero expectation value on the product state Eq.~\eqref{eq:prod-state},  as $\expval{\delta E_{pq}}{\Psi_X} = 0$.
Removing this term (which we will later treat perturbatively) we obtain the mean-field interaction
\begin{equation}
    H_\text{mf} = \sum_{pq\in A}\sum_{rs\in B} g^{\prime}_{pqrs} \left[E_{pq}\gamma^B_{rs} + \gamma^A_{pq}E_{rs} - \gamma^A_{pq} \gamma^B_{rs} \right].
\end{equation}
We can finally define $H^0$ as
\begin{align}\label{eq:ham_zero}
    H^0 &= H_A + H_B + H_\text{mf},
\end{align}
where all terms are operators with support on only a single fragment, thus the ground state $\ket{\Psi^0}$ of $H^0$ is a product state of the form Eq.~\eqref{eq:prod-state}.
All the terms we removed from $H$ to construct $H^0$ have zero expectation value on $\ket{\Psi^0}$, thus it is the \textit{lowest energy product state that respects the on-fragment symmetries.}

\topic{algorithm for finding $\Psi^0$}
To find $\ket{\Psi^0}$ we minimize $E_0 = \expval{H^0}$ by self-consistently solving separate ground state problems on each fragment. 
Consider the decomposition
\begin{align}\label{eq:e0_decomp}
    E_0 &=
    E_A + E_B + E_\text{mf},
\end{align}
where $E_X = \mel{\Psi_X}{H_X}{\Psi_X}$ can be evaluated on a single fragment $X$ and $E_\text{mf} = \sum_{pq\in A}\sum_{rs\in B} g^\prime_{pqrs} \gamma^A_{pq} \gamma^B_{rs}$ is the mean-field inter-fragment coupling depending on the fragment 1-RDMs.
To find $\ket{\Psi_A}$ and $\ket{\Psi_B}$, we iteratively solve for the ground state of the following coupled Hamiltonians:
\begin{align}
    H^{\rm eff}_A &= H_A + \sum_{pq\in A}\sum_{rs\in B} g^{\prime}_{pqrs} E_{pq}\gamma^B_{rs}\label{eq:ham_eff_A}\\
    H^{\rm eff}_B &= H_B + \sum_{pq\in A}\sum_{rs\in B} g^{\prime}_{pqrs} \gamma^A_{pq}E_{rs}\label{eq:ham_eff_B},
\end{align}
thus minimizing all the terms Eq.~\eqref{eq:e0_decomp}.
We outline the whole procedure in Algorithm~\ref{alg:psi_0_solver}.
Note that this algorithm can be readily generalized to other MC solvers within the fragment that provide access to the state RDMs (e.g. the variational quantum eigensolver, discussed in Sec.~\ref{sec:quantum-algorithms}).
\begin{algorithm}[t]
    \caption{Fragment embedding \label{alg:psi_0_solver}}
    
    \KwIn{%
    Active space integrals $h_{pq}$, $g_{pqrs}$. 
    Fragment-partitioned sets of orbitals: occupied $(O_\text{A}, O_\text{B})$, and virtual $(V_\text{A}, V_\text{B})$ subspaces.
    Convergence threshold $\tau$.
    }
    
    \KwOut{Optimal product state $\ket{\Psi^0} = \ket{\Psi_A}\ket{\Psi_B}$}
    
    \vspace{1em}

    Start from a Hartree-Fock state: $\gamma^B_{pq} \gets 2\delta_{pq}$ for $p, q \in O_B$\;
    $\Delta E,\, E^0 \gets \infty$\;

    \While{
        $\Delta E > \tau$}{
        $\ket{\Psi_A},  E^{\rm eff}_A \gets$ ground state of $H^{\rm eff}_A$ of Eq.~\eqref{eq:ham_eff_A} \;
        $\gamma^A_{pq} \gets \mel{\Psi_A}{E_{pq}}{\Psi_A}$, $p, q \in A$\;
        $\ket{\Psi_B}, E^{\rm eff}_B\gets$ ground state of $H^{\rm eff}_B$ of Eq.~\eqref{eq:ham_eff_B}\;
        $\gamma^B_{rs} \gets \mel{\Psi_B}{E_{rs}}{\Psi_B}$, $p, q \in B$\;
        $E_\text{mf} \gets \sum_{pq\in A}\sum_{rs\in B} g^\prime_{pqrs} \gamma^A_{pq} \gamma^B_{rs}$\;
        $\Delta E \gets E^0 - (E^{\rm eff}_A + E^{\rm eff}_B - E_\text{mf})$\;
        $E^0 \gets (E^{\rm eff}_A + E^{\rm eff}_B - E_\text{mf})$
    }
    
    \Return $\ket{\Psi^0}=\ket{\Psi_A}\ket{\Psi_B}$
    
\end{algorithm}

\newcolumntype{A}{>{}l<{\hspace{.5em}}}
\newcolumntype{B}{>{\hspace{.5em}}l<{\hspace{.5em}}}
\newcolumntype{C}{>{\hspace{.5em}}l<{}}
\begin{table*}[t]\centering
    \begin{tabularx}{\textwidth}{A B B C}
    \hline \\[-1.5ex]
      &
      \textbf{Perturbation} 
      &
      \textbf{Perturbing functions} $\left\{\ket{\Psi_{\mu\nu}}\right\}$ 
      & 
      \textbf{Fragment matrix element}
    \\[1.5ex] \hline \\[1.5ex] 
    1. &
    ${
        \begin{aligned}[t]
            H^\prime_{\rm disp}= 
            &\sum_{pq\in A}\sum_{rs\in B} g^{\prime}_{pqrs} \left(E_{pq} - \gamma^A_{pq}\right) \left(E_{rs} - \gamma^B_{rs}\right)
        \end{aligned}
    }$
    & 
    ${
        \begin{aligned}[t]
            E&_{tu}E_{vw}\ket{\Psi^0} \\
            &\begin{bmatrix}
                tu\in A, \, vw \in B
            \end{bmatrix}
        \end{aligned}
    }$
    &
    ${
        \mel{\Psi_X}{E_{lk} H^{\rm eff}_{X} E_{tu}}{\Psi_X}
    }$
    \\[1.5ex] \\ 
    2. 
    &
    ${
        \begin{aligned}[t]
            H^\prime_{\rm 1CT} = &\sum_{p\in A}\sum_{q\in B} \left[ h_{pq}  - \sum_{r\in A}g_{prrq}\right] E_{pq}\\
            &+\sum_{p \in B}\sum_{q \in A} \left[ h_{pq}  - \sum_{r\in B}g_{prrq}\right] E_{pq}\\
            &+\sum_{pqr \in A}\sum_{s \in B} g_{pqrs} E_{pq}\left[E_{rs} + E_{sr}\right]\\
            &+\sum_{pqr \in B}\sum_{s \in A} g_{pqrs} E_{pq}\left[E_{rs} + E_{sr}\right]
        \end{aligned}
    }$
    &
    ${
        \begin{aligned}[t]
            E&_{tu}E_{vw}\ket{\Psi^0} \\
            &\begin{bmatrix}
             tuv \in A, w \in B\\
             tuw \in A, v \in B\\
             tuv \in B, w \in A\\
             tuw \in B, v \in A
            \end{bmatrix}
        \end{aligned}
    }$
    &
    ${
        \begin{aligned}[t]
            \mel{\Psi_X}{a_{m}E_{lk} H^{\rm eff}_{X} E_{tu} a^\dagger_{v}}{\Psi_X}\\
            \mel{\Psi_X}{a^\dagger_{m}E_{lk} H^{\rm eff}_{X} E_{tu} a_{v}}{\Psi_X}
        \end{aligned}
    }$
    \\[1.5ex] \\ 
    3. 
    &
    ${
        \begin{aligned}[t]
            H^\prime_{\rm 2CT} = &\frac{1}{2}\sum_{pr \in A}\sum_{qs \in B}g_{pqrs}E_{pq}E_{rs} \\
            &+ \frac{1}{2}\sum_{pr \in B}\sum_{qs \in A}g_{pqrs}E_{pq}E_{rs}
        \end{aligned}
    }$
    & 
    ${
        \begin{aligned}[t]
            E&_{tu}E_{vw}\ket{\Psi^0} \\
            &\begin{bmatrix}
                tv \in A, uw \in B\\
                uw \in A, tv \in B\\
            \end{bmatrix}
        \end{aligned}
    }$ 
    &
    ${
        \begin{aligned}[t]
            \mel{\Psi_X}{a_{l}a_{k} H^{\rm eff}_{X} a^\dagger_{t} a^\dagger_{u}}{\Psi_X}\\
            \mel{\Psi_X}{a^\dagger_{l}a^\dagger_{k} H^{\rm eff}_{X} a_{t} a_{u}}{\Psi_X}
        \end{aligned}
    }$
    \\[1.5ex] \\ 
    4. &
    ${
        \begin{aligned}[t]
            H^\prime_{\rm TT} = -\sum_{pq \in A}\sum_{rs\in B} g_{psrq} t_{pq, rs}
        \end{aligned}
    }$ 
    & 
    ${
        \begin{aligned}
            &t_{tu, vw}\ket{\Psi^0} \\ 
            &\begin{bmatrix}
                tu\in A, \, vw \in B
            \end{bmatrix}
        \end{aligned} 
    }$
    &
    ${
        \begin{aligned}[t]
            &\mel{\Psi_X}{T^{(1,0)}_{l k} H^{\rm eff}_{X} T^{(1,0)}_{tu}}{\Psi_X}\\
            &\mel{\Psi_X}{T^{(1,1)}_{l k} H^{\rm eff}_{X} T^{(1,-1)}_{tu}}{\Psi_X}\\
            &\mel{\Psi_X}{T^{(1,-1)}_{l k} H^{\rm eff}_{X} T^{(1,1)}_{tu}}{\Psi_X}
        \end{aligned}
    }$
    \vspace{1.5ex}
    \\
    \hline
\end{tabularx}
\setcounter{table}{0}
\caption{\label{tab:pert_cost}
    \textbf{Summary of the perturbations and the cost of PT2.}
    We summarize here the perturbing functions and cost for each of the perturbations. 
    The rightmost column reports the form of the matrix elements of $H^0$ required to compute each perturbation; estimating these matrix elements on the fragment state is the most expensive part of FragPT2.
    If done naively by writing out the full fragment Hamiltonians as a contraction between integrals and  this could require estimating 4-RDMs for the dispersion, double-charge transfer (2CT) and triplet-triplet (TT) perturbations, and 5-RDMs for the single-charge transfer (1CT) perturbation.
}
\end{table*}

\subsection{Multi-reference perturbation theory}\label{sec:pt2}

\topic{Inter-fragment interactions}
While the $\ket{\Psi^0}$ retrieved from Algorithm~\ref{alg:psi_0_solver} is a solid starting point, it neglects the correlations between the fragments.
If the fragments are sufficiently separated, we expect these correlations to be minimal and recoverable by perturbation theory.
We propose using second-order perturbation theory to retrieve the correlation energy of these interactions. 
The inter-fragment interaction terms can be classified in four categories, based on whether they conserve charge and/or total spin on each fragment: dispersion $H^\prime_\text{disp}$ (which conserves both charge and spin of the fragments), single-charge transfer $H^\prime_\text{1CT}$ and double-charge transfer $H^\prime_\text{2CT}$ (that conserve charge nor spin), and triplet-triplet coupling $H^\prime_\text{TT}$ (that conserves charge but not local spin). 
Thus, the complete decomposition of the Hamiltonian reads:
\begin{align}\label{eq:ham_decomp}
    H = H^0 + H^\prime_{\rm disp} + H^\prime_{\rm 1CT} + H^\prime_{\rm 2CT} + H^\prime_{\rm TT}.
\end{align}
The definition of these terms is given in Table~\ref{tab:pert_cost} and their derivation is reported in Appendix~\ref{app:ham_decomp}.
We will treat the different perturbations in Eq.~\eqref{eq:ham_decomp} one at a time.
First notice that for every perturbation in Eq.~\eqref{eq:ham_decomp}, the first order energy correction is zero: $E^1 = \mel{\Psi^0}{H^\prime}{\Psi^0} = 0$. We will focus solely on the second order correction to the energy.

\topic{perturbing functions}
To proceed, we need to choose a basis of \textit{perturbing functions} $\left\{\ket{\Psi_\mu}\right\}$ used to define the first-order correction to the wavefunction
\begin{align}\label{eq:firstorder_wfn}
    \ket{\Psi^1} = \sum_\mu C_\mu \ket{\Psi_\mu}.
\end{align}
For the exact second order perturbation energy, we should consider all Slater determinants that can be obtained by applying the terms within $H$ to the set of reference determinants. 
While this full space of perturbing functions is smaller than the complete eigenbasis of $H$, it is still unpractically large, and approximations need to be introduced. To choose a compact and expressive basis, we look at the perturbation under consideration.
Every perturbative Hamiltonian can be expanded in a linear combination of two-body operators:
\begin{align}\label{eq:pert_ham}
    H^\prime = \sum_{\mu\in A}\sum_{\nu\in B} g_{\mu\nu} O^A_{\mu} O^B_{\nu}.
\end{align}
where $O^X_{\mu}$ is either identity or a product of Fermionic operators on fragment $X$ and $g_{\mu \nu}$ are combinations of one- and two-electron integrals (see Table~\ref{tab:pert_cost} for their explicit form).
Consider the following (non-orthogonal) basis:
\begin{align}\label{eq:pert_func}
    \ket{\Psi_{\mu \nu}} = O^A_{\mu}O^B_{\nu} \ket{\Psi^0}.
\end{align}
This partially-contracted basis is a natural choice for compactly representing the wavefunctions that interact with $\ket{\Psi^0}$ through the perturbations in $H'$ \cite{angeli2001introduction}.

\topic{Solve linear equations}
Following the choice of perturbing functions, we estimate the matrix elements $\mel**{\Psi_{\mu\nu}}{H^0}{\Psi_{\kappa\lambda}}$ in this basis.
The overlap $\bra{\Psi_{\mu\nu}}\ket{\Psi_{\kappa\lambda}}$
must also be computed in order to be able to contract with the $g_{\mu\nu}$ to yield $\mel**{\Psi^0}{H^{\prime}}{\Psi_{\mu\nu}}$.
To obtain the coefficients $C_{\mu\nu}$ that define the first-order correction to the wavefunction, we solve the following linear equations:
\begin{align}\label{eq:le_pt2}
    \sum_{\kappa\lambda}\mel**{\Psi_{\mu\nu}}{H^0 - E^0}{\Psi_{\kappa\lambda}} C_{\kappa\lambda} + \mel**{\Psi_{\mu\nu}}{H^\prime}{\Psi^0} = 0.
\end{align}
Then the second-order correction to the energy is given by:
\begin{align}\label{eq:energy_pt2}
    E^2 = \mel{\Psi^0}{H^\prime}{\Psi^1}= \sum_{\mu\nu} \mel**{\Psi^0}{H^\prime}{\Psi_{\mu\nu}} C_{\mu\nu}.
\end{align}
The total second-order PT correction can be expressed as the sum of the different perturbations:
\begin{align}\label{eq:total_e2}
    E^2 = E^2_{\rm disp} + E^2_{\rm 1CT} + E^2_{\rm 2CT} + E^2_{\rm TT}.
\end{align}
The procedure is summarized in Algorithm~\ref{alg:pt2}.

\begin{algorithm}[h]
    \caption{FragPT2}
    \label{alg:pt2}
    
    \KwIn{%
    Active space integrals and orbital partitioning as in Algorithm~\ref{alg:psi_0_solver}.
    Zeroth-order product state $\ket{\Psi^0} = \ket{\Psi_A}\ket{\Psi_B}$.
    }
    
    \KwOut{Second-order perturbative corrections: $E^2_{\rm disp},\, E^2_{\rm 1CT}, \, E^2_{\rm 2CT}$ and $E^2_{\rm TT}$}

    \vspace{1em}

    \For{$H^\prime \in \{H^\prime_{\rm disp},\,H^\prime_{\rm 1CT},\,H^\prime_{\rm 2CT},\,H^\prime_{\rm TT}\}$}{
        Choose perturbing functions as in Eq.~\eqref{eq:pert_func}\;
        Compute matrix elements $\mel**{\Psi_{\mu\nu}}{H^0}{\Psi_{\kappa\lambda}}$\;
        Compute matrix elements $\mel**{\Psi_{\mu\nu}}{H^\prime}{\Psi^0}$\;
        $C_{\mu \nu} \gets$ solution of Eq.~\eqref{eq:le_pt2}\;
        $E^2 \gets \sum_{\mu\nu} \mel**{\Psi^0}{H^\prime}{\Psi_{\mu\nu}} C_{\mu\nu}$
    }
    
    \Return $E^2_{\rm disp},\, E^2_{\rm 1CT}, \, E^2_{\rm 2CT}$ and $E^2_{\rm TT}$
    
\end{algorithm}

\topic{Cost of matrix elements}
Computing the matrix elements $\mel**{\Psi_{\mu\nu}}{H^0}{\Psi_{\kappa\lambda}}$ is the most expensive part of our algorithm. The tensor product form of the zeroth-order wavefunction significantly reduces the algorithm's cost by allowing the matrices to factorize in the expectation values of operators on the different fragments, that in turn can be expressed as combinations of fragment RDMs.
We outline the idea here, and refer the reader to Appendix~\ref{app:frag_mel} for the formal derivation for every perturbation:
\begin{align}\label{eq:fragmented_mel}
   \mel**{\Psi_{\mu\nu}}{H^0}{\Psi_{\kappa\lambda}} \hspace{-2em}&\hspace{2em}= \\ \nonumber
    & \mel**{\Psi_A}{{O^A_{\mu}}^\dagger H^{\rm eff}_A O^A_{\kappa}}{\Psi_A} \mel**{\Psi_B}{{O^B_{\nu}}^\dagger O^B_{\lambda}}{\Psi_B} \\ \nonumber
    +& \mel**{\Psi_A}{{O^A_{\mu}}^\dagger O^A_{\kappa}}{\Psi_A} \mel**{\Psi_B}{{O^B_{\nu}}^\dagger H^{\rm eff}_B  O^B_{\lambda}}{\Psi_B} \\ \nonumber
    +& E_\text{mf} \mel**{\Psi_A}{{O^A_{\mu}}^\dagger O^A_{\kappa}}{\Psi_A} \mel**{\Psi_B}{{O^B_{\nu}}^\dagger O^B_{\lambda}}{\Psi_B}. 
\end{align}
If there are $N_{A}N_{B}$ two-body terms in Eq.~\eqref{eq:pert_ham}, the number of matrix elements that one needs to estimate on each fragment is $\frac{1}{2}N_{A}N_{B}(N_{A}N_{B} + 1)$.
However, if the amount of matrix elements becomes too expensive,
it is possible to alleviate the cost without sacrificing much accuracy, for example by using a more compact basis of perturbing functions.
For a discussion of further reductions of the cost, see Section~\ref{sec:efficiency}.

\section{Numerical demonstration}
\label{sec:numerics}
In this section we demonstrate our method by applying it to a range of molecular systems that are well-suited targets for bipartite fragmentation. We have chosen three sets of systems.
The first system consists of two \ce{N2} molecules at a distance of $2.0\si{\angstrom}$, with a (close to equilibrium) bond length of $1.2\si{\angstrom}$.
In contrast to the other structures, we do not need to cut through a covalent bond and can treat each molecule as a separate fragment. 
We examine the results of our method while stretching the nitrogen bond in one of the fragments; this is known to rapidly increase static correlation in this system and thus is a good benchmark for the multi-reference method.
The second type of systems we consider comprises a set of aromatic dimers, where two aromatic rings of different kinds are connected by a single covalent bond. Cutting through this bond, we investigate the correlation energies of the dimers with respect to the dihedral angle of the ring alignment. 
These systems exhibit strong correlation whenever the rings are in the same plane and low correlation when the rings are perpendicular to each other: they are thereby suitable to benchmark both regimes. 
The final system is butadiene, as the simplest example of the class of polyene molecules that are much studied as 1-D model systems \cite{yarkony1977band,ghosh2008orbital} as well as for their importance in various applications \cite{marder1997large,christensen2004linear,barrett2019recent}.
Here we cut through the single covalent bond between the middle carbons and investigate the correlation energy with respect to the stretching of the double bonds in a single fragment.
This system, albeit slightly artificial, is intriguing due to the significant static correlation within the fragments induced by the dissociating bonds, coupled with substantial dynamic inter-fragment correlation.

\subsection{Numerical simulation details}
We construct the localized orbitals using a localization scheme implemented in the ROSE code \cite{rose}. 
The FragPT2 method is implemented completely inside the quantum chemical open-source software package PySCF \cite{sun2020recent}. Algorithm~\ref{alg:psi_0_solver} uses the FCI solver of the program to get the optimal product state of the fragments. 
The matrix elements in Eq.~\eqref{eq:fragmented_mel} by exploiting the software capabilities to manipulate CI-vectors and estimate higher order RDMs. 
Finally we implemented Algorithm~\ref{alg:pt2} that solves Eq.~\eqref{eq:le_pt2} and \eqref{eq:energy_pt2} for every perturbation in Eq.~\eqref{eq:ham_decomp}. To assess the accuracy of our algorithm, we compare the fragment embedding energy $E^0$ (from Algorithm \ref{alg:psi_0_solver}), the FragPT2 energy $E^0 + E^2$ including the perturbative correction (from Algorithm \ref{alg:pt2}), and the exact ground state energy $E^{\rm exact}$ of the Hamiltonian in Eq.~\eqref{eq:ham_tot} (calculated with CASCI in an a full-molecule active space of dobule size).
The \ce{N2} dimer and aromatic dimer calculations are done in a cc-pVDZ basis set, while butadiene is treated in a 6-31G basis.

\subsection{\ce{N2} dimer}

\begin{figure}[t]
    \centering
    \includegraphics[width=\linewidth]{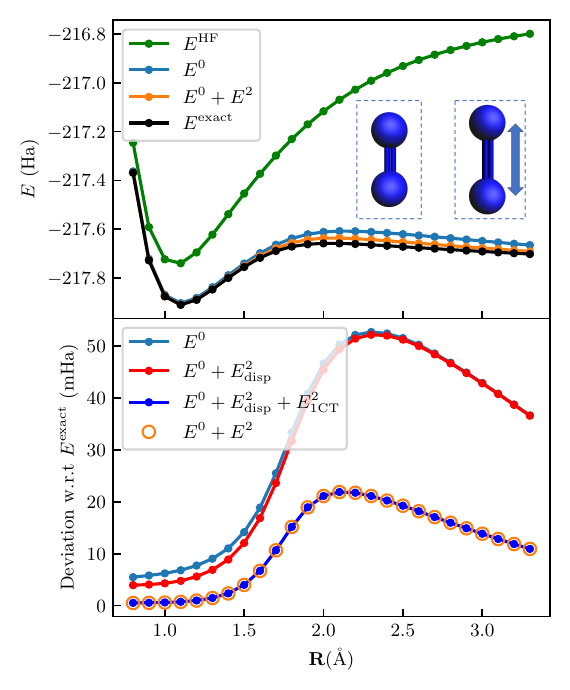}
    \caption{\textbf{Potential energy curve for the \ce{N2} dimer.}
    The upper panel shows a comparison of the curves obtained through Hartree-Fock ($E^\text{HF}$), fragment embedding ($E^0$), FragPT2 ($E^0 + E^2$), and full-molecule CASCI (exact).
    The two \ce{N2} are parallel and at a distance of \si{2\angstrom}, and the bond distance of the right dimer is varied.
    The fragment active spaces each comprise six electrons in six spatial orbitals, corresponding to the triple bonding and anti-bonding orbitals. 
    Hartree-Fock performs poorly due to the strong intra-frament correlation.
    The fragment embedding energy $E^0$ captures the correct behaviour of the system, while $E^2$ gives an additional, small correction in the direction of the exact solution.
    The bottom panel reports the deviation w.r.t.~the exact result over the pogential energy curve, where we sequentially add the different perturbative corrections described in Table~\ref{tab:pert_cost}. 
    We first add the dispersion correction $E^2_\text{disp}$ (red line) and then the single-charge transfer contribution $E^2_\text{1CT}$ (blue line), showing the other contributions are zero by plotting the full FragPT2 energy $E^0 + E^2$ (orange dots).
    }
    \label{fig:n2_dimer}
\end{figure}

As an initial test system, we consider a dimer of nitrogen molecules, i.e. \ce{N2-N2}. To increase the static correlation within the fragment, we dissociate one of the nitrogen molecules. This bond breaking is modeled using three occupied and three virtual localized orbitals in the active space, representing the $\sigma$ bond and the two $\pi$ bonds. This results in an active space of six electrons in six orbitals for each fragment. The results in Figure~\ref{fig:n2_dimer} clearly demonstrate the failure of the Hartree-Fock method due to the high degree of correlation within the fragment. Our multi-reference solver within the localized active spaces successfully addresses this issue, with $E^0$ providing a good description of the ground state. There is some minor inter-fragment correlation, and our perturbative correction brings us closer to the exact solution. 

Our data further shows that the perturbative correction arises mainly from the single-charge transfer contribution.
Notably, the double-charge transfer and triplet-triplet coupling are zero everywhere.
Additionally, we find that for stretched bond lengths, the dispersion interaction between the fragments is minimal.
The ability to identify the character of the relevant interactions between fragments is a further advantage of our method. 

\begin{figure*}[t]
    \centering
    \includegraphics[width=\textwidth]{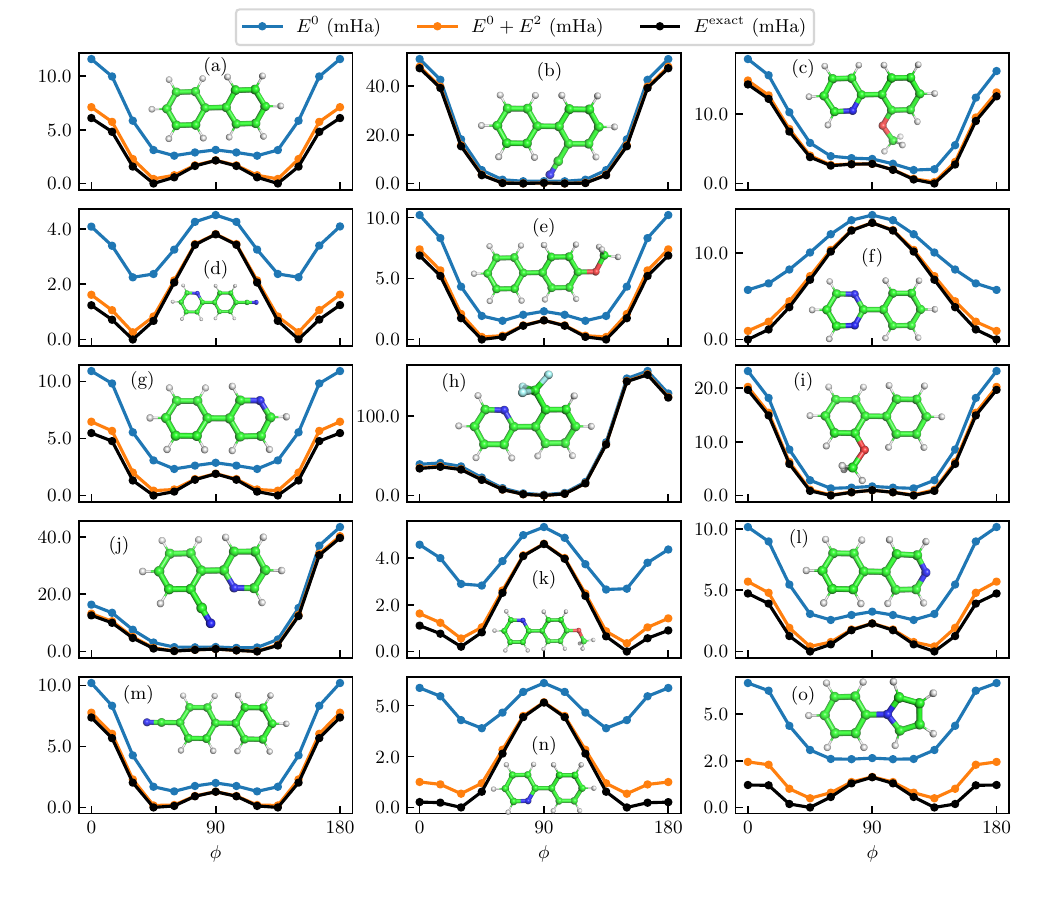}
    \caption{\textbf{Relative potential energy curves for the set of aromatic dimers}, where we vary the dihedral angle $\phi$ of the two dimers. The molecular orbitals are localized on the fragments naturally defined by the two aromatic rings (including the respective ligands). In principle, for each dimer, we select the active space of six electrons in six orbitals on each fragment that comprise the conjugated $\pi-\pi^*$ system (with some exceptions elaborated on in Section~\ref{sec:aromatic_dimers}).
    Thus our method cuts down the space for the exact calculation (twelve electrons in twelve orbitals) into half. This is small enough to verify our method against an exact CASCI calculation. The blue line represents the fragment embedding energy $E^0$. 
    The orange line includes the second-order perturbation energy for all considered perturbations, representing the FragPT2 energy $E^0 + E^2$. 
    The black line reports the exact calculation $E^{\rm exact}$.
    All reported energies are relative to the minimum of $E^{\rm exact}$.
    The considered  molecules are, in row-first order: (a) biphenyl (b) 2-cyanobiphenyl (c) 2-(2-methoxyphenyl)pyridine (d) 2-(4-cyanophenyl)pyridine (e) 4-methoxybiphenyl (f) 2-phenylpyrimidine (g) 3-phenylpyridine (h) 2-(2-trifluoromethylphenyl)pyridine (i) 2-methoxybiphenyl (j) 2-(2-cyanophenyl)pyridine (k) 2-(4-methoxyphenyl)pyridine (l) 4-phenylpyridine (m)4-cyanobiphenyl (n) 2-phenylpyridine (o) N-phenylpyrrole.
    \label{fig:aromatic_dimer_abs}}
\end{figure*}

\subsection{Aromatic dimers}\label{sec:aromatic_dimers}

Here we focus on aromatic dimers, i.e. molecules with two aromatic rings that are attached by a single covalent bond. The simplest such system considered is two phenyl rings, known as biphenyl, shown in Figure~\ref{fig:active_space}. As the biphenyl case is highly symmetric, other similar molecules can be generated by substituting various ligands for one of the hydrogen atoms, or a nitrogen for a carbon in the phenyl rings. In this manner, we generate a comprehensive benchmark on a variety of systems. 
Our set of examples is motivated from the different classes of biaryl systems studied by Sanfeliciano \textit{et al.}~in the context of drug design  \cite{sanfeliciano2018}.

To construct the fragment active spaces, we consider the conjugated $\pi-\pi^*$ system on each ring, typically resulting in six electrons distributed across six orbitals for each fragment. There are a few exceptions to this rule. For pyrrole rings, the relevant aromatic orbitals comprise six electrons in five orbitals.
Furthermore, for rings that include a \ce{CN} or \ce{OCH3} substituent [i.e.~(c-f), (i-k) and (m) in Figure~\ref{fig:aromatic_dimer_abs}], there is a low-lying $\pi$ orbital and  high-lying $\pi^*$ orbital that mix with a $p$ orbital of the substituent. These orbitals are excluded from the active space of these fragments, reducing the active space to four electrons in four orbitals. This only provides additional insight into the performance of our method with asymmetric active space sizes in the fragments.

For each dimer, we vary the dihedral angle $\phi$ of the two planes spanned by the rings, thus rotating over the covalent bond. This gives a potential energy curve with a high variance of correlation energy: if the rings are perpendicular, the aromatic systems are localized and the correlation between the fragments is low. Instead, if the rings are aligned, we expect to see a high amount of correlation between the fragments, and thus a breakdown of the description of $E^0$. The results of our method compared to the exact energies are given in Figure~\ref{fig:aromatic_dimer_abs}.

\begin{figure}[t]
    \centering
    \includegraphics{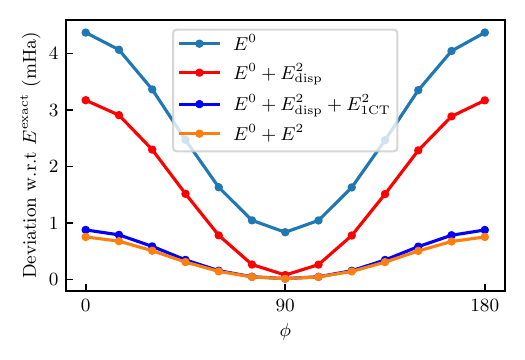}
    \caption{
    \textbf{Average errors for the aromatic dimer set.}
    Mean deviation in total energy with respect to the exact result for the complete set of aromatic dimers shown in Figure~\ref{fig:aromatic_dimer_abs}, where we vary the dihedral angle $\phi$ of the two aromatic rings. 
    We show the result of sequentially adding the different perturbative corrections described in Table~\ref{tab:pert_cost}. 
    The top curve represents the error of fragment embedding energy $E^0$. 
    We first add the dispersion correction $E^2_\text{disp}$, which is giving a constant shift along the dihedral angle. 
    Then, the (single) charge transfer correction $E^2_\text{1CT}$ crucially corrects for the behaviour where the rings are aligned. 
    Finally, we add the double-charge transfer term $E^2_\text{2CT}$ and triplet-triplet term $E^2_\text{TT}$  together, recovering the final FragPT2 energy $E^0 + E^2$. These last terms contribute an additional small shift to the aligned rings configuration.
    }
    \label{fig:MAD_contrib}
\end{figure}

Our data shows that, for each of the molecules and values of $\phi$, $E^0$ recovers at least $93\%$ of the correlation energy (with an average of $97\%$). 
While this is high in absolute terms,
the shape of the potential energy curves for these models can be qualitatively wrong. As expected, a product state is not a good approximation if the rings are aligned, as the aromatic system will be delocalized over the molecule. This causes the interactions between the fragments to play a more significant role. The product state is on the other hand a good approximation when the rings are perpendicular, there pushing $E^0$ to $99\%$ of the correlation energy. This causes an imbalance between the two configurations and calls for the need to treat the interactions. When we compute the second-order perturbation energy $E^2$, it is shown in Figure~\ref{fig:aromatic_dimer_abs} that sometimes $E^0$ finds a different minimum than the exact state. In these cases especially, the perturbative corrections need to be calculated to give a more correct shape of the potential energy curve. In Figure~\ref{fig:MAD_contrib}, one can see that the division of the perturbation energies can be very constructive in determining the important contributions of the system in question. In case of aromatic dimers, two interactions are important: dispersion and single-charge transfer. While the former takes care of a constant shift over the dihedral angles, the latter is much larger when the aromatic rings are aligned, thus crucial in retrieving the right behaviour of PES. The double-charge transfer and triplet-triplet spin exchange terms are not important in these class of molecules.

\subsection{Butadiene}

Butadiene (\ce{C4H6}) is the final test system that we consider. 
We define the two fragments by cutting through the middle bond of the molecule. 
We study the energy of the system while stretch the double bond onto dissociation inside one of the fragments, thereby testing our method to increasing amounts of static correlation inside the fragment. 
Dissociating the bond additionally causes the leftover molecule to be a radical, thus increasing significantly the strength of the interaction between the fragments. 

We define the active spaces by taking the $\pi-\pi^*$ and $\sigma-\sigma^*$ system of the double bonds of both fragments.
This results in an active space of four electrons in four orbitals for each fragment.

\begin{figure}[t]
    \centering
    \includegraphics[width=\linewidth]{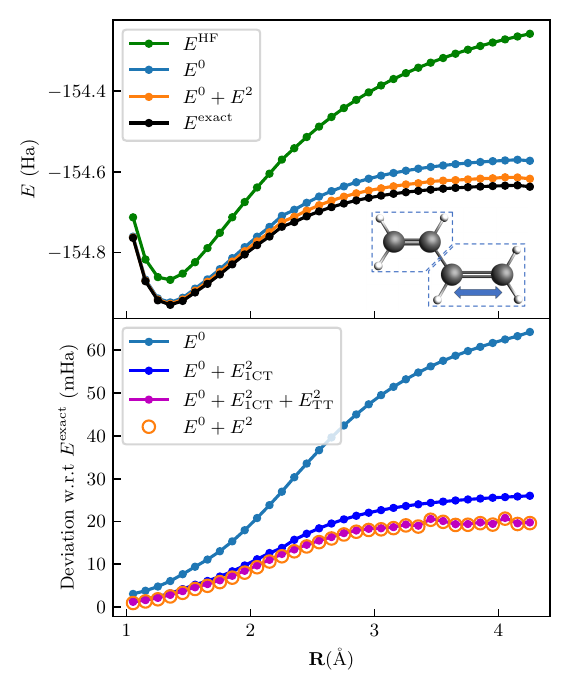}
    \caption{
    \textbf{Potential energy curves of butadiene.} 
    The fragments are chosen by cutting through the middle bond and subsequently stretching the double bond of one of the fragments, as illustrated in the inset.
    The curves are color coded like in Figure~\ref{fig:n2_dimer}, and show that both intra-fragment and inter-fragment correlations are important to recover the correct behaviour.
    In particular, inter-fragment correlations are explained by the vicinity of the two fragments and by the radical that is left over after dissociation of the stretched bond.
    In the lower panel, we show once more the result of sequentially adding the different perturbative corrections described in Table~\ref{tab:pert_cost}. We first add the single-charge transfer contribution $E^2_\text{1CT}$ (blue line) and then the triplet-triplet coupling $E^2_\text{TT}$ (purple line), showing the other contributions are zero by plotting the full FragPT2 energy $E^0 + E^2$ (orange dots).
    }
    \label{fig:buta_energy}
\end{figure}

The potential energy curves are shown in the upper panel of Figure~\ref{fig:buta_energy}. 
It can be clearly seen that the multi-reference product state is a correct description at the equilibrium geometry, but its performance is somewhere in between the Hartree-Fock and the exact solution at dissociation. 
To improve on it, we clearly need the perturbative corrections.

If we analyze the contributions to the perturbative correction plotted in the lower panel of Figure~\ref{fig:buta_energy}, we see that $H^\prime_{\rm 1CT}$ interaction is the most important (contributing around $80\sim88\%$ to $E^2$).
Notably, in this system the $H^\prime_{\rm TT}$ contribution is large as well (contributing around $8\sim12\%$ to $E^2$). This is in line with chemical intuition, as this system has low-lying triplet states \cite{mosher1973triplet}; a singlet-coupled double triplet excitation may therefore contribute significantly to the ground state wave function.
Again, the ability to separately analyse the different classes of inter-fragment interactions is useful here, as it allows to consider the correlation in polyenes in terms of products of local excitations.

\section{Outlook}\label{sec:outlook}

\subsection{Computational efficiency}
\label{sec:efficiency}
In order to estimate the perturbative corrections in Table~\ref{tab:pert_cost} we have to construct high-order $k$-RDMs for all $k \leq 5$.
These RDMs are tensors with up to 10 indices: constructing and storing them explicitly is computationally expensive.
Several methods to evaluate and store high order RDMs in a compressed form have been proposed in the context of e.g.~NEVPT2 theory \cite{guo2021approximations1,guo2021approximations2}.
Resolution of identity (RI) \cite{eichkorn1995auxiliary}, cumulant expansions \cite{zgid2009study}, tensor contraction with integrals \cite{kollmar2021efficient,chatterjee2020extended} are some of the ways to circumvent this bottleneck. 
Future resesarch should cosider how these methods can be applied in the specific case of FragPT2.

To further improve the efficiency of FragPT2, we can 
consider modifications to the part our algorithm that calculates the perturbative corrections.
For example, to circumvent the need to calculate all the different elements of the RDMs, we can compress the basis of perturbing functions in Eq.~\eqref{eq:pert_func}. 
One option is to set the coefficients $C_\mu$ of Eq.~\eqref{eq:firstorder_wfn} by the integrals of the perturbation $H'$ under consideration.
This gives a single (unnormalized) perturbing function $\ket{\Psi^1} = H' \ket{\Psi^0}$, known in literature as a \textit{strongly contracted basis}. 
This strongly contracted form has applied with some success in the context of NEVPT2~\cite{angeli2001introduction}.  
The bottleneck of the algorithm then becomes estimating higher order powers of the Hamiltonian and the perturbations on $\ket{\Psi^0}$, effectively equivalent to the first order of a moment expansion~\cite{cioslowski1987, angeli2001introduction}.
Another possible approach to reducing the cost of computing perturbations relies on stochastic formulations of MRPT, which have also been studied in the context of strongly-contracted NEVPT2 \cite{mahajan2019multireference,blunt2020efficient,anderson2020efficient}.
In these approaches, the necessary quantities were determined in a quantum Monte-Carlo (QMC) framework.

\subsection{Integration with quantum algorithms}\label{sec:quantum-algorithms}

In this manuscript, we focused on solving the single fragments with FCI, but our framework is compatible with any method that can recover RDMs of fragment wavefunctions. 
Quantum algorithms have emerged as promising tools for tackling classically-hard electronic structure problems, but they come with specific limitations distinct from those of classical algorithms. 
Fragmentation and embedding techniques are critical for defining tasks suited to quantum algorithms, enabling a focus on strongly-correlated active sites while reducing problem sizes. 
Recent studies have explored integrating quantum algorithms into embedding schemes, including SAPT (for both near-term variational \cite{malone2022simulation,loipersberger2022interaction} and fault-tolerant \cite{cortes2023faulttolerant} approaches) and LASSCF \cite{otten2022localized,mitraLocalized2024}. 
In this context, we discuss integrating FragPT2 with the variational quantum eigensolver (VQE) \cite{peruzzo2014,mcclean2016theory}.


The VQE prescribes to prepare on a quantum device an ansatz state $\ket{\Psi(\boldsymbol\theta)}$, as a function of a set of classical parameters $\boldsymbol\theta$ which are then optimized to minimize the state energy $E(\boldsymbol\theta) = \bra{\Psi(\boldsymbol\theta)} H \ket{\Psi(\boldsymbol\theta)}$.
Having access to a quantum device allows to produce states which can be hard to represent on a classical computer, enabling the implementation of ans\"atze such as unitary coupled cluster \cite{bartlettAlternative1989,peruzzo2014} and other heuristic constructions \cite{grimsley2019adaptive,anselmetti2021local,cerezo2021}; however, sampling the energy and other properties from the quantum state incurs a large sampling cost, which is worsened by the required optimization overhead.

Integrating FragPT2 with the VQE is straightforward. 
For each fragment $X$, a separate parameterized wavefunction $\ket{\Psi_X(\boldsymbol\theta_X)}$ is represented, reconstructing an ansatz $\ket{\Psi_A(\boldsymbol\theta_A)}\ket{\Psi_A(\boldsymbol\theta_B)}$ for the product state Eq.~\eqref{eq:prod-state}.
As no quantum correlation is needed, multiple wavefunctions can be prepared in parallel in separate quantum devices, or even serially on the same device; this can allow to treat larger chemical systems with limited-size quantum devices.
We can find the lowest-energy product state directly by minimizing the expectation value of the Hamiltonian
\begin{equation}
    E(\boldsymbol\theta_A, \boldsymbol\theta_B) = \bra{\Psi_A(\boldsymbol\theta_A)}\bra{\Psi_B(\boldsymbol\theta_B)}
    H
    \ket{\Psi_A(\boldsymbol\theta_A)}\ket{\Psi_B(\boldsymbol\theta_B)}.
\end{equation}
this energy can be estimated by measuring the one- and two-body reduced density matrices separately on each fragment.
As shown in Section~\ref{sec:ham_decomp}, the minimum energy product state matches the solution of the mean field embedding.

Integrating VQE with fragmentation techniques can help describe binding energies, proposed in literature with a method based on symmetry adapted perturbation theory and termed SAPT(VQE) \cite{malone2022simulation, loipersberger2022interaction}. SAPT(VQE) addresses the same terms as Algorithm~\ref{alg:psi_0_solver}, but uses a perturbative expansion instead of mean-field coupling for terms dependent on fragment 1- and 2-RDMs. It employs two non-orthogonal orbital sets for the fragments, limiting this method to non-covalently bonded fragments. Inspired by SAPT(VQE), Algorithm~\ref{alg:psi_0_solver} could be adapted to measure interaction energies. A thorough comparison of the two methods and studying their dependence on molecular orbitals and atomic basis set is a promising area for future work. SAPT has also recently been applied to fault-tolerant quantum algorithms, overcoming some of the limitations of near-term devices \cite{cortes2023faulttolerant}.

As per Algorithm~\ref{alg:pt2}, to recover the perturbative corrections accounting for inter-fragment interactions we need to extract higher-order reduced density matrices from each fragment's wavefunction.
Perturbation theory for the variational quantum eigensolver has been studied in the context of recovering dynamical correlations \cite{tammaro2023nelectron,krompiec2022}.
Using measurement optimization techniques from \cite{bonet-monroig2020,huggins2021}, estimating all the elements of the $k$-RDMs on a fragment active space of $N$ orbitals to a precision $\epsilon$ requires $O(\epsilon^{-2} N^k)$ samples.
In practice this makes naively estimating the perturbative corrections very costly, especially for the single-charge transfer terms $H'_\text{1CT}$ that require $5$-RDMs (see Table~\ref{tab:pert_cost}).
An interesting direction for future research might consider using shadow tomography and its fermionic extension \cite{huang2020,wan2023matchgate} to estimate RDMs to all orders at the same time.

\subsection{Further extensions}\label{sec:extensions}

\emph{Extension to multiple fragments ---}
This paper focused on the case of two active fragments. However, it is relatively straightforwardly applied to more. The lowest energy product state can be retrieved by trivially extending the algorithm, looping through the fragments and solving exactly the active fragment feeling the mean-field of the inactive fragments, until reaching convergence. Subsequently, we can treat the inter-fragment interactions that can span four fragments at a time at most (as the Hamiltonian is a two-body operator), which is a coupled charge transfer excitation. While the perturbing functions then have to be extended to these types of excitations, the matrix elements that one has to estimate will factorize in the same way, and the algorithm will not be more costly than for two fragments (i.e. no higher order RDMs will have to be estimated). Working out the exact expressions and implementing a truly many-fragment algorithm is part of future work.

\emph{Localized orbitals beyond Hartree-Fock ---}
The Hartree-Fock determinant is known to be an unstable reference in dissociating systems and other highly-correlated molecules \cite{helgaker2000, szabo1996}.
To generate the input orbitals, one might want to change from a cheap mean-field method to a slightly more expensive CASSCF calculation with a small active space. As the localization scheme can handle any input orbitals, our method can be trivially adapted to a better choice of reference orbitals that already takes into account some correlation. 
Additionally, one can include intra-fragment orbital-optimization during the fragment embedding (Algorithm~\ref{alg:psi_0_solver}). A simple approach would involve using a CASSCF solver on the individual fragments, with orbital rotations constrained to each fragment to keep the fragments separated. This could enhance the method's accuracy and provide a better starting point for perturbation theory.
In this spirit, a version LASSCF \cite{hermes2019multiconfigurational} or vLASSCF \cite{hermes2020variational} could be recovered as an extension of our method where orbital rotations between fragment active spaces are also allowed and optimized self-consistently.

\emph{NEVPT2-like perturbations ---}
So far we have treated the interactions only inside the complete active space, i.e. our $H$ from Eq.~\eqref{eq:ham_tot} involves indices within either active fragment. To retrieve more of the dynamical correlation energy, the core idea of NEVPT2 is to include excitations involving also the inactive orbitals in a perturbative way. We can build on top of our previous approach by including additional perturbations and perturbing functions.
Correspondingly, we can augment our Hamiltonian from Eq.~\eqref{eq:ham_tot} as,
\begin{align}
    H = H^{0} + H^{'}_{act} + H^{'}_{inact}
\end{align}
where $H^{'}_{inact}$ consists of the various classes of perturbations involving excitations from the core to the active space, the active space to the virtual space and the core to the active space. For the form of these perturbations, see reference~\citenum{angeli2001introduction}. It is straightforward to extend the methods from NEVPT2 to the case of multiple active fragments, and again the matrix elements will factorize on different fragments in the same way, relieving the need to estimate additional matrix elements on the multi-reference fragment solvers.

\section{Conclusion}\label{sec:conclusion}

In this work, we introduced a novel multi-reference multi-fragment embedding framework called FragPT2.
We showed that our method gives accurate results for a reduced cost in active space size, especially when the fragments are well-separated.
Our comprehensive numerical benchmarks on a variety of molecules show that: 1. Intra-fragment static correlation can be retrieved by an MC product state ansatz ($E^0$) 2. Inter-fragment correlation can be treated as a perturbative correction ($E^2$) 3. A combination of these is needed to recover the correct shape of the potential energy curve.
Using a decomposition of the Hamiltonian based on fragment symmetries, we can distinguish the contributions to the inter-fragment correlation in $E^2$, providing insight into important interactions within the studied systems.
Furthermore, our adapted localization scheme allows to define molecular fragments that cut through covalent bonds. 
In this case, perturbative corrections describing inter-fragment charge transfer (and, to a lesser extent, triplet-triplet spin exchange) are crucial for accurately describing the system.
Future research directions include improving the efficiency of high-order RDM estimation, integrating FragPT2 with variational quantum algorithms, and extensions to multiple fragments for broader applicability. 

Our multi-reference embedding scheme could find broad applications, for instance in understanding the spatial dependence of the correlation energy in $\pi$-stacked systems and other biochemically important systems \cite{phipps2015energy}, modelling supramolecular complex formation \cite{albelda2012supramolecular} in metal ion separation, or in analyzing metallophylic interactions \cite{sculfort2011intramolecular,pyykko1994predicted,otero2014metallophilic}.

\section*{Acknowledgements}
We thank Dr. Arno F{\"o}rster, Sarathchandra Khandavilli and Detlef Hohl for stimulating discussions.
We thank Seenivasan Hariharan and Matthias Degroote for useful feedback.
We acknowledge support from Shell Global Solutions BV.

\newpage
\bibliography{main}

\clearpage
\onecolumngrid

\appendix
\section{Hamiltonian decomposition}\label{app:ham_decomp}
In this appendix, we give a detailed derivation of the terms in the decomposition Eq.~\eqref{eq:ham_decomp} of the full active-space molecular Hamiltonian. 

We start from the full Hamiltonian Eq.~\eqref{eq:ham_tot}, and we rewrite the quartic excitation operators $e_{pqrs} = E_{pq}E_{rs} - \delta_{qr}E_{ps}$ in terms of the quadratic $E_{pq}$, obtaining
\begin{align}\label{eq:ham_full_expanded}
    H = \sum_{pq} h_{pq} E_{pq} - \frac{1}{2}\sum_{pq} \sum_{r} g_{prrq} E_{pq} + \frac{1}{2}\sum_{pqrs} g_{pqrs} E_{pq}E_{rs}.
\end{align}
We will separately deal with the terms that conserve charge on each fragment (in Appendix~\ref{app:ham_decomp_cc}) and those that transfer charge between fragments (in Appendix~\ref{app:ham_decomp_ct}).
It can be easily identified whether a term preserves charge on each fragment by counting the number of electrons moved across orbitals, as all orbitals $\{p, q, r, s\}$ pertain to either fragment $A$ or $B$.

\subsection{Charge-conserving terms}\label{app:ham_decomp_cc}

In this section we derive the charge-conserving inter-fragment terms $H'_\text{disp}$ and $H'_\text{TT}$, as well as the on-fragment Hamiltonians $H_A$ and $H_B$.

The one-body term of Eq.~\eqref{eq:ham_full_expanded} only conserves charge if $p$ and $q$ are both in the $A$ fragment or both in the $B$ fragment, these terms will respectively be part of $H_A$ and $H_B$.
The two-body term of Eq.~\eqref{eq:ham_full_expanded} includes terms where all $p,q,r,s$ are part of the same fragment: these will also be part of $H_A$ and $H_B$. 

The other possible options that preserve charge while including terms on both fragments are:
\begin{align}\label{eq:ham_split_2body}
\begin{split}
    &\frac{1}{2}\sum_{pq\in A}\sum_{rs \in B} g_{pqrs} E_{pq}E_{rs}  + \frac{1}{2} \sum_{pq\in B}\sum_{rs \in A} g_{pqrs} E_{pq}E_{rs}\\
    +&\frac{1}{2}\sum_{pq \in A}\sum_{rs \in B} g_{psrq} E_{ps}E_{rq} + \frac{1}{2} \sum_{rs \in B} \sum_{pq \in A} g_{rqps} E_{rq}E_{ps}.
\end{split}
\end{align}
It is straightforward to show that the first two terms are equivalent by using the symmetry $g_{pqrs} = g_{rspq}$ and the excitation operator commutation relations $\left[E_{pq},E_{rs}\right] = E_{ps}\delta_{qr} - E_{rq}\delta_{ps}$. 
These terms represent the Coulomb interactions between the fragments. 
The latter two terms describe the exchange interactions between the fragments.

We rewrite the exchange term using Fermionic commutation rules -- using the notation $p_X$ being an orbital index on fragment $X$ we get
\begin{align}
\begin{split}\label{eq:exchange_operator}
    E_{p_As_B}E_{r_Bq_A} =& \sum_{\sigma \tau} a^\dagger_{p_A \sigma} a_{s_B \sigma} a^\dagger_{r_B \tau} a_{q_A \tau}\\
    =& \sum_{\sigma \tau} \left(\delta_{r_Bs_B}\delta_{\sigma \tau}a^\dagger_{p_A \sigma} a_{q_A \tau}-a^\dagger_{p_A \sigma} a_{q_A \tau} a^\dagger_{r_B \tau} a_{s_B \sigma}\right)\\
    = & \delta_{r_Bs_B} E_{p_Aq_A} \\
    - & a^\dagger_{p_A \alpha} a_{q_A \alpha} a^\dagger_{r_B \alpha} a_{s_B \alpha}
    - a^\dagger_{p_A \beta} a_{q_A \beta} a^\dagger_{r_B \beta} a_{s_B \beta}\\
    - &a^\dagger_{p_A \alpha} a_{q_A \beta} a^\dagger_{r_B \beta} a_{s_B \alpha}
    -a^\dagger_{p_A \beta} a_{q_A \alpha} a^\dagger_{r_B \alpha} a_{s_B \beta}\\
    = & \delta_{r_Bs_B} E_{p_Aq_A} 
    - S^{0,0}_{p_Aq_A}S^{0,0}_{r_Bs_B}\\
    - &T^{1,0}_{p_Aq_A}T^{1,0}_{r_Bs_B} +
    T^{1,1}_{p_Aq_A}T^{1,-1}_{r_Bs_B} +
    T^{1,-1}_{p_Aq_A}T^{1,1}_{r_Bs_B}
\end{split}
\end{align}
where we use the definition of the spin operators \cite{helgaker2000}:
\begin{align}
S^{(0,0)}_{p_Xq_X} &= \frac{1}{\sqrt{2}}E_{p_Xq_X} = \frac{1}{\sqrt{2}}(a^\dagger_{p_X \alpha} a_{q_X \alpha} + a^\dagger_{p_X \beta} a_{q_X \beta}) \label{eq:s00}\\
T^{(1,0)}_{p_Xq_X} &= \frac{1}{\sqrt{2}}(a^\dagger_{p_X \alpha} a_{q_X \alpha} - a^\dagger_{p_X \beta} a_{q_X \beta}) \label{eq:t00}\\
T^{(1,1)}_{p_Xq_X} &= - a^\dagger_{p_X \alpha} a_{q_X \beta} \label{eq:t11}\\
T^{(1,-1)}_{p_Xq_X} &= a^\dagger_{p_X \beta} a_{q_X \alpha} \label{eq:t1m1}
\end{align}
and
\begin{align}
\begin{split}
a^\dagger_{p \alpha} a_{q \alpha} = \frac{1}{\sqrt{2}}(S^{(0,0)}_{pq} + T^{(1,0)}_{pq})\\
a^\dagger_{p \beta} a_{q \beta} = \frac{1}{\sqrt{2}}(S^{(0,0)}_{pq} - T^{(1,0)}_{pq}).
\end{split}
\end{align}

Notice that the last three terms of Eq.~\eqref{eq:exchange_operator} conserve the total spin of the system, but flip the local spin of the individual fragments. 
Separating out these terms from the expansion of Eq.~\eqref{eq:ham_split_2body} we obtain the triplet-triplet interaction Hamiltonian:
\begin{align}
    H^\prime_{\rm TT} = 
    \sum_{pq \in A} \sum_{rs \in B} g_{psrq} 
    T^{1,0}_{p_Aq_A}T^{1,0}_{r_Bs_B} +
    T^{1,1}_{p_Aq_A}T^{1,-1}_{r_Bs_B} +
    T^{1,-1}_{p_Aq_A}T^{1,1}_{r_Bs_B} 
    = -\sum_{pq \in A}\sum_{rs\in B} g_{psrq} t_{pq, rs}
\end{align}
where $t_{pq, rs}=
T^{1,0}_{pq}T^{1,0}_{rs} -
T^{1,1}_{pq}T^{1,-1}_{rs} -
T^{1,-1}_{pq}T^{1,1}_{rs}$.

After substracting this term from Eq.~\eqref{eq:ham_split_2body}, we arrive at the following expression for the Hamiltonian that includes all fragment charge-conserving and spin-conserving terms: $H_A + H_B + H_{AB}$, which can be easily split in the three terms
\begin{align}
    H_A = &\sum_{pq \in A} h_{pq} E_{pq} - \frac{1}{2} \sum_{pqr \in A} g_{prrq} E_{pq} + \sum_{pqrs \in A} g_{pqrs} E_{pq}E_{rs}\\
    H_B = &\sum_{pq \in B} h_{pq} E_{pq} - \frac{1}{2} \sum_{pqr \in B} g_{prrq} E_{pq} + \sum_{pqrs \in B} g_{pqrs} E_{pq}E_{rs}\\
    H_\text{mf} + H'_\text{disp} = &\sum_{pq\in A}\sum_{rs \in B} \left[g_{pqrs} - \frac{1}{2} g_{psrq}\right] E_{pq}E_{rs},
\end{align}
where the last row can be further split in a mean-field interaction term $H_\text{mf}$ and a dispersion term $H'_\text{disp}$: the mean-field interaction is defined self-consistently on the basis of the solution of the Hamitlonian $H^0 = H_A + H_B + H_\text{mf}$, as we showed in Section~\ref{sec:ham_decomp}.

\subsection{Charge transfer terms}\label{app:ham_decomp_ct}

We now work on separating the terms that involve charge transfers between the fragments. 
As the molecular Hamiltonian contains only one-body and two-body terms, we only need to consider single-charge and double-charge transfers, respectively classified as part of $H'_{1CT}$ and $H'_{2CT}$.

We first isolate the single-charge transfer terms. 
These include the single-body terms of Eq.~\eqref{eq:ham_full_expanded} where $p$ and $q$ pertain to different fragments:
\begin{align} \label{eq:1ct-one_body_term}
    \sum_{p \in A}\sum_{q \in B} h_{pq} E_{pq} + \sum_{p \in B}\sum_{q \in A} h_{pq} E_{pq};
\end{align}
along with the two-body terms where three indices pertain to a fragment and one pertains to the other:
\begin{align}
\begin{split}
    &\frac{1}{2}\left[ \sum_{pqr \in A}\sum_{s\in B} + \sum_{pqs \in A}\sum_{r\in B} + \sum_{prs \in A}\sum_{q\in B} + \sum_{qrs \in A}\sum_{p\in B}  \right] \Big(g_{pqrs} \left(E_{pq}E_{rs} - \delta_{qr} E_{ps}\right)\Big)\\
    +&\frac{1}{2}\left[ \sum_{pqr \in B}\sum_{s\in A} + \sum_{pqs \in B}\sum_{r\in A} + \sum_{prs \in B}\sum_{q\in A} + \sum_{qrs \in B}\sum_{p\in A}  \right] \Big( g_{pqrs} \left(E_{pq}E_{rs} - \delta_{qr} E_{ps}\right) \Big),    
\end{split}
\end{align}
where for brevity we write multiple sums (in brackets) sharing the same summand (in parentheses).
We can simplify this using the symmetries of the two-body integral $g_{pqrs} = g_{rspq} = g_{qprs} = g_{pqsr}$ and the commutation relations of excitation operators $\left[E_{pq},E_{rs}\right] = E_{ps}\delta_{qr} - E_{rq}\delta_{ps}$, obtaining
\begin{align}
\begin{split}
    &-\sum_{p\in A}\sum_{q\in B} \left[\sum_{r\in A}g_{prrq}\right] E_{pq}
            -\sum_{p \in B}\sum_{q \in A} \left(\sum_{r\in B}g_{prrq}\right) E_{pq}\\
            &+\sum_{pqr \in A}\sum_{s \in B} g_{pqrs} E_{pq}\left(E_{rs} + E_{sr}\right)
            +\sum_{pqr \in B}\sum_{s \in A} g_{pqrs} E_{pq}\left(E_{rs} + E_{sr}\right).
\end{split}
\end{align}
Combining this with the one-body term Eq.~\eqref{eq:1ct-one_body_term} we define the single-charge transfer term
\begin{align}
\begin{split}
    H^\prime_{\rm 1CT} = &\sum_{p\in A}\sum_{q\in B} \left[ h_{pq}  - \sum_{r\in A}g_{prrq}\right] E_{pq}
            +\sum_{p \in B}\sum_{q \in A} \left[ h_{pq}  - \sum_{r\in B}g_{prrq}\right] E_{pq}\\
            &+\sum_{pqr \in A}\sum_{s \in B} g_{pqrs} E_{pq}\left[E_{rs} + E_{sr}\right]
            +\sum_{pqr \in B}\sum_{s \in A} g_{pqrs} E_{pq}\left[E_{rs} + E_{sr}\right].
\end{split}
\end{align}

The double-charge transfer is simpler, as there are just two two-body terms that allow for a double-charge transfer:
\begin{align}
    H^\prime_{\rm 2CT} = \frac{1}{2}\sum_{pr \in A}\sum_{qs \in B}g_{pqrs}E_{pq}E_{rs} 
            + \frac{1}{2}\sum_{pr \in B}\sum_{qs \in A}g_{pqrs}E_{pq}E_{rs}.
\end{align}

One can easily verify that $H = H_A + H_B + H_\text{mf} + H'_\text{disp} + H'_\text{TT} + H'_\text{1CT} + H'_\text{2CT}$.

\section{Fragment matrix elements}\label{app:frag_mel}
This section contains derivations of the expressions for the zeroth-order Hamiltonian matrix elements $\mel{\Psi_{\mu \nu}}{H^0}{\Psi_{\kappa \lambda}}$ for every perturbation $H^\prime = \sum_{\mu \nu} g_{\mu \nu} O^A_{\mu} O^B_{\nu}$. We will focus here on estimating these matrix elements exactly without any approximation, resulting in the need for higher order RDMs. For a discussion of future work to improve efficiency, see Section~\ref{sec:efficiency}.
In general, the expressions are given as:
\begin{align}\label{eq:app_fragment_mel}
\begin{split}
   \mel**{\Psi_{\mu\nu}}{H^0}{\Psi_{\kappa\lambda}} &= 
     \mel**{\Psi_A}{{O^A_{\mu}}^\dagger H^{\rm eff}_A O^A_{\kappa}}{\Psi_A} \mel**{\Psi_B}{{O^B_{\nu}}^\dagger O^B_{\lambda}}{\Psi_B} 
    + \mel**{\Psi_A}{{O^A_{\mu}}^\dagger O^A_{\kappa}}{\Psi_A} \mel**{\Psi_B}{{O^B_{\nu}}^\dagger H^{\rm eff}_B  O^B_{\lambda}}{\Psi_B}\\
    &\qquad\qquad\qquad\qquad\qquad\qquad\qquad\qquad\qquad+ E_\text{mf} \mel**{\Psi_A}{{O^A_{\mu}}^\dagger O^A_{\kappa}}{\Psi_A} \mel**{\Psi_B}{{O^B_{\nu}}^\dagger O^B_{\lambda}}{\Psi_B},
\end{split}\\
     \mel**{\Psi^0}{H^\prime}{\Psi_{\kappa \lambda }} &= \sum_{\mu \nu} g_{\mu \nu}\mel**{\Psi_A}{{O^A_{\mu}}^\dagger O^A_{\kappa}}{\Psi_A} \mel**{\Psi_B}{{O^B_{\nu}}^\dagger O^B_{\lambda}}{\Psi_B},
\end{align}
where we defined the perturbing functions as $\ket{\Psi_{\mu\nu}} = O^A_{\mu}O^B_\nu \ket{\Psi_A}\ket{\Psi_B}$. We have:
\begin{align}\label{eq:frag_heff_mel}
\begin{split}
    \mel**{\Psi_X}{O^X_{\mu}H^{\rm eff}_X O^X_{\nu}}{\Psi_X} =& \sum_{pq \in X} \left( h_{pq} + \sum_{rs \in Y}g^\prime_{pqrs} \gamma^Y_{rs}  - \sum_{r \in X} g_{prrq}\right)  \mel**{\Psi_X}{O^X_{\mu} E_{pq} O^X_{\nu}}{\Psi_X} \\
    &+ \sum_{pqrs \in X} g_{pqrs} \mel**{\Psi_X}{O^X_{\mu} E_{pq}E_{rs} O^X_{\nu}}{\Psi_X}
\end{split}
\end{align}
Thus, the relevant operator matrix elements to estimate are $\mel**{\Psi_X}{O^X_{\mu} E_{pq}E_{rs} O^X_{\nu}}{\Psi_X}$, $\mel**{\Psi_X}{O^X_{\mu} E_{pq} O^X_{\nu}}{\Psi_X}$ and $\mel**{\Psi_X}{O^X_{\mu} O^X_{\nu}}{\Psi_X}$.

\subsection{Dispersion}
For the case of the dispersion perturbation, we use the perturbing functions 
\begin{align}
    E_{tu}E_{vw}\ket{\Psi^0} 
    \begin{bmatrix}
        tu\in A, \, vw \in B
    \end{bmatrix}.
\end{align}
Thus, we straightforwardly identify $O^X_\mu \rightarrow E_{pq}$ and the most expensive matrix element to compute is:
\begin{align}
    \mel**{\Psi_X}{E_{vw} E_{pq}E_{rs} E_{tu}}{\Psi_X},
\end{align}
i.e. a 4-particle reduced density matrix.

\subsection{Single-charge transfer}
The single-charge transfer case is a bit more complicated. We like to preserve the total spin, so we use the spin-free excitation operators to define the perturbing functions as:
\begin{align}\label{eq:pert_func_1ct}
    E&_{tu}E_{vw}\ket{\Psi^0} \quad
    \begin{bmatrix}
     tuv \in A, w \in B\\
     tuw \in A, v \in B\\
     tuv \in B, w \in A\\
     tuw \in B, v \in A
    \end{bmatrix}.
\end{align}
That means a straightforward decomposition into $O^A_\mu O^B_\nu$ is more intricate because of the sum over spin. Instead, we have $\sum_{\sigma} E_{p_X q_X} a^\dagger_{v_X \sigma} a_{w_Y \sigma}$ and $\sum_{\sigma} E_{p_Y q_Y} a^\dagger_{v_X \sigma} a_{w_Y \sigma}$. Let us work out the matrix elements explicitly for the first case in Eq.~\eqref{eq:pert_func_1ct}. This is easily generalizable to the other cases. 
The total matrix element becomes:
\begin{align}
\begin{split}
   \mel**{\Psi_{k_A l_A m_A n_B}}{H^0}{\Psi_{t_A u_A v_A w_B}} = 
     \sum_{\sigma \tau}\mel**{\Psi_A}{a_{m\sigma}E_{lk} H^{\rm eff}_A E_{tu} a^\dagger_{v\tau}}{\Psi_A} \mel**{\Psi_B}{a^\dagger_{n\sigma} a_{w \tau}}{\Psi_B} \\
     + \sum_{\sigma \tau}\mel**{\Psi_A}{a_{m\sigma}E_{lk}E_{tu} a^\dagger_{v\tau}}{\Psi_A} \mel**{\Psi_B}{a^\dagger_{n\sigma} H^{\rm eff}_B a_{w \tau}}{\Psi_B} + E_\text{mf} \sum_{\sigma \tau}\mel**{\Psi_A}{a_{m\sigma}E_{lk}E_{tu} a^\dagger_{v\tau}}{\Psi_A} \mel**{\Psi_B}{a^\dagger_{n\sigma} a_{w \tau}}{\Psi_B}
\end{split}
\end{align}
As $E_{pq}$ and $H^{\rm eff}_X$ preserve spin, and $\ket{\Psi_X}$ are eigenfunctions of $S^2$, we can replace the double sum over spin by a single one as $\sigma = \tau$.
Now substituting $O^X_\mu \rightarrow E_{tu}a^\dagger_{v\sigma}$ in Eq.~\eqref{eq:frag_heff_mel} the most expensive object to estimate will be:
\begin{align}
    \mel**{\Psi_X}{a_{m\sigma}E_{lk} E_{pq} E_{rs} E_{tu} a^\dagger_{v\sigma}}{\Psi_X},
\end{align}
i.e. a 5-particle reduced density matrix.

\subsection{Double-charge transfer}
For the double-charge transfer we have the following set of perturbing functions:
\begin{align}
    E&_{tu}E_{vw}\ket{\Psi^0} 
            \begin{bmatrix}
                tv \in A, uw \in B\\
                uw \in A, tv \in B\\
            \end{bmatrix}.
\end{align}
This makes the total matrix element equal to:
\begin{align}
\begin{split}
   \mel**{\Psi_{k_A l_B m_A n_B}}{H^0}{\Psi_{t_A u_B v_A w_B}} =&
     \sum_{\sigma \tau \kappa \lambda}\mel**{\Psi_A}{a_{m\sigma}a_{k\tau} H^{\rm eff}_A a^\dagger_{t\kappa} a^\dagger_{v\lambda}}{\Psi_A} \mel**{\Psi_B}{a^\dagger_{n\sigma} a^\dagger_{l\tau} a_{u \kappa} a_{w \lambda}}{\Psi_B} \\
     &+ \sum_{\sigma \tau \kappa \lambda}\mel**{\Psi_A}{a_{m\sigma}a_{k\tau} a^\dagger_{t\kappa} a^\dagger_{v\lambda}}{\Psi_A} \mel**{\Psi_B}{a^\dagger_{n\sigma} a^\dagger_{l\tau}  H^{\rm eff}_B a_{u \kappa} a_{w \lambda}}{\Psi_B} \\
     &+ E_\text{mf} \sum_{\sigma \tau \kappa \lambda}\mel**{\Psi_A}{a_{m\sigma}a_{k\tau} a^\dagger_{t\kappa} a^\dagger_{v\lambda}}{\Psi_A} \mel**{\Psi_B}{a^\dagger_{n\sigma} a^\dagger_{l\tau} a_{u \kappa} a_{w \lambda}}{\Psi_B}.
\end{split}
\end{align}
Here there also simplifications possible regarding the sum over spin. Namely, only the following options are non-zero: $\sigma, \tau, \kappa, \lambda \in \left\{ \alpha\alpha\alpha\alpha, \beta\beta\beta\beta, \alpha\beta\alpha\beta, \alpha\beta\beta\alpha, \beta\alpha\alpha\beta, \beta\alpha\beta\alpha \right\}$. Regardless, making the identification $O^X_\mu \rightarrow a^\dagger_{t\sigma} a^\dagger_{v\sigma}$ in Eq.~\eqref{eq:frag_heff_mel}, the most expensive object to estimate is:
\begin{align}
    \mel**{\Psi_X}{a_{m\sigma}a_{k\tau} E_{pq} E_{rs} a^\dagger_{t\kappa} a^\dagger_{v\lambda}}{\Psi_X},
\end{align}
and similarly for $O^X_\mu \rightarrow a_{t\sigma} a_{v\sigma}$, this is as expensive as estimating a 4-particle reduced density matrix.

\subsection{Triplet-triplet}
The triplet-triplet perturbing functions are given by:
\begin{align}
    t_{tuvw}\ket{\Psi^0} 
    \begin{bmatrix}
        tu\in A, \, vw \in B
    \end{bmatrix},
\end{align}
where $t_{tuvw} = T^{1,0}_{tu}T^{1,0}_{vw} -
    T^{1,1}_{tu}T^{1,-1}_{vw} -
    T^{1,-1}_{tu}T^{1,1}_{vw}$ (see Appendix~\ref{app:ham_decomp_cc} for their definition).
Observe that, for any fragment operator $O^X$ that preserves spin, all matrix elements $\mel**{\Psi_X}{T^{(1,m^\prime)} O^X T^{(1,m^\prime)}}{\Psi_X}$ are \textit{only} non-zero if $m + m^\prime = 0$, where $m, m^\prime \in \{-1,0,-1\}$.
Thus, we can make the following statement:
\begin{align}
\begin{split}
    &\mel**{\Psi^0_{klmn}}{O^A O^B}{\Psi^0_{tuvw}}  \\
    = &\mel{\Psi^0}{t_{k_A l_A, m_B n_B}^\dagger O^A O^B t_{t_A u_A,v_B w_B }}{\Psi^0}\\
    = &\mel{\Psi_A}{T^{(1,0)}_{l k} O^A  T^{(1,0)}_{tu}}{\Psi_A} \mel{\Psi_B}{T^{(1,0)}_{n m} O^B T^{(1,0)}_{vw}}{\Psi_B}  \\
    + &\mel{\Psi_A}{T^{(1,1)}_{l k} O^A  T^{(1,-1)}_{tu}}{\Psi_A} \mel{\Psi_B}{T^{(1,-1)}_{n m} O^B T^{(1,1)}_{vw}}{\Psi_B} \\
    + &\mel{\Psi_A}{T^{(1,-1)}_{l k} O^A  T^{(1,1)}_{tu}}{\Psi_A} \mel{\Psi_B}{T^{(1,1)}_{n m} O^B T^{(1,-1)}_{vw}}{\Psi_B}
\end{split}
\end{align}
Finally, the matrix element of $H^0$ in this basis is thus equal to:
\begin{align}
\begin{split}
   \mel**{\Psi_{k_A l_A m_B n_B}}{H^0}{\Psi_{t_A u_A v_B w_B}} =&
     \sum_{m + m^\prime = 0}\mel**{\Psi_A}{T^{(1,m)}_{lk} H^{\rm eff}_A T^{(1, m^\prime)}_{tu}}{\Psi_A} \mel**{\Psi_B}{T^{(1, m^\prime)}_{nm} T^{(1,m)}_{vw}}{\Psi_B} \\
     &+ \sum_{m + m^\prime = 0}\mel**{\Psi_A}{T^{(1,m)}_{lk} T^{(1, m^\prime)}_{tu}}{\Psi_A} \mel**{\Psi_B}{T^{(1, m^\prime)}_{nm} H^{\rm eff}_A T^{(1,m)}_{vw}}{\Psi_B} \\
     &+ E_\text{mf} \sum_{m + m^\prime = 0}\mel**{\Psi_A}{T^{(1,m)}_{lk} T^{(1, m^\prime)}_{tu}}{\Psi_A} \mel**{\Psi_B}{T^{(1, m^\prime)}_{nm} T^{(1,m)}_{vw}}{\Psi_B}.
\end{split}
\end{align}
The most expensive object to estimate in the triplet-triplet case is then, identifying $O^X_\mu \rightarrow T^{(1,m)}_{tu}$ in Eq.~\eqref{eq:frag_heff_mel}:
\begin{align}
    \mel**{\Psi_X}{T^{(1,m)}_{lk} E_{pq} E_{rs} T^{(1,m^\prime)}_{tu}}{\Psi_X},
\end{align}
where $m+m^\prime = 0$. This is equivalent in cost to measuring a 4-particle reduced density matrix.

\end{document}